\documentclass[aps,prc,reprint,superscriptaddress,nofootinbib,floatfix]{revtex4-1}

\usepackage{graphicx}
\usepackage{multirow}
\usepackage{color}
\usepackage{epsfig}
\usepackage{amssymb}
\usepackage{bbm}
\usepackage{amsmath}
\usepackage{color}
\usepackage{lineno}
\usepackage{enumitem}
\usepackage{scalerel,stackengine}
\usepackage{comment}
\usepackage[normalem]{ulem}
\usepackage{subfigure}

\def\nuc#1#2{\relax\ifmmode{}^{#1}{\protect\text{#2}}\else${}^{#1}$#2\fi}

\newcommand{\be}{\begin{eqnarray}}
\newcommand{\ee}{\end{eqnarray}}

\newcommand{\bwt}{\begin{widetext}}
\newcommand{\ewt}{\end{widetext}}

\stackMath
\newcommand\reallywidehat[1]{%
\savestack{\tmpbox}{\stretchto{%
  \scaleto{%
    \scalerel*[\widthof{\ensuremath{#1}}]{\kern-.6pt\bigwedge\kern-.6pt}%
    {\rule[-\textheight/2]{1ex}{\textheight}}
  }{\textheight}%
}{0.5ex}}%
\stackon[1pt]{#1}{\tmpbox}%
}

\begin{document}

\title{Kolmogorov-Arnold networks in nuclear binding energy prediction}



\author{Hao Liu}


\affiliation{School of Physics Science and Engineering, Tongji University, Shanghai 200092, China.}

\author{Jin Lei}
\email[Contact author: ]{jinl@tongji.edu.cn}

\affiliation{School of Physics Science and Engineering, Tongji University, Shanghai 200092, China.}

\author{Zhongzhou Ren}

\affiliation{School of Physics Science and Engineering, Tongji University, Shanghai 200092, China.}

\begin{abstract}
This study explores the application of Kolmogorov-Arnold networks (KANs) in predicting nuclear binding energies, leveraging their ability to decompose complex multiparameter systems into simpler univariate functions. By utilizing data from the Atomic Mass Evaluation (AME2020) and incorporating features such as atomic number, neutron number, and shell effects, KANs achieved a significant lower root mean square error (0.26~MeV), surpassing traditional models. The symbolic regression analysis yielded simplified analytical expressions for binding energies, aligning with classical models like the liquid drop model and the Bethe-Weizsäcker formula. These results highlight KANs' potential in enhancing the interpretability and understanding of nuclear phenomena, paving the way for future applications in nuclear physics and beyond.
\end{abstract}


\date{\today}%
\maketitle

\section{Introduction}\label{sec:intro}
The atomic nucleus, a quintessential quantum many-body system, exhibits remarkable structural complexity~\cite{Hammer20}. Binding energies (BE), synonymous with nuclear mass, are key characteristics of atomic nuclei and play an indispensable role in elucidating various nuclear phenomena. These phenomena include nuclear shapes, shell effects, pairing effects, and the emergence and disappearance of magic numbers~\cite{weizsacker1935theorie,PhysRevLett.114.202501,PhysRevC.90.024320,RevModPhys.75.1021}. Furthermore, binding energies are crucial in the synthesis of superheavy elements and in understanding nuclear astrophysical processes like the r-process~\cite{ARNOULD200797} and x-ray bursts~\cite{xraySchatz_2017}. Therefore, both theoretical predictions and experimental measurements of nuclear binding energies are essential for advancements in nuclear physics.

Over the years, numerous theories and methods have been developed for predicting nuclear binding energies~\cite{BETHE1936,weizsacker1935theorie,MYERS1996141,mean-field,hadizadeh2020three,Hammer20,RevModPhys.81.1773,ann2024,PhysRevC.105.064306,PhysRevC.106.014305,PhysRevC.104.014315,PhysRevC.98.034318,PhysRevC.106.L021303,yuan2024reliable,ml2024,munoz2024discoveringnuclearmodelssymbolic}. These include the Bethe-Weizsäcker formula~\cite{BETHE1936,weizsacker1935theorie}, the Thomas-Fermi model~\cite{MYERS1996141}, the Hartree-Fock-Bogoliubov mean field model~\cite{mean-field}, and ab initio methods~\cite{hadizadeh2020three,Hammer20,RevModPhys.81.1773}. These physics-based models have achieved commendable results within their respective applicable ranges. Concurrently, data-driven methods such as multilayer perceptrons~\cite{ann2024,PhysRevC.105.064306,PhysRevC.106.014305,PhysRevC.104.014315,PhysRevC.98.034318,PhysRevC.106.L021303}, Gaussian processes~\cite{ml2024,yuan2024reliable}, and support vector machines~\cite{ml2024} have also found successful applications in the prediction of atomic nucleus masses.

However, traditional physical models necessitate a profound understanding of the inherent mechanisms of nuclear physical systems and face challenges when managing complex relationships. On the other hand, while data-driven machine learning methods can handle complex non-linear relationships, they often require substantial data and computational resources. In the data-rich field of nuclear physics, identifying intricate relationships among variables is a formidable challenge. This predicament underscores the need for innovative approaches to handle the complexity of nuclear data and to extract meaningful insights.

Kolmogorov-Arnold networks (KANs) offer a promising solution~\cite{liu2025kan}. KANs provide a simplified yet robust approach by decomposing complex multi-parameter systems into manageable univariate functions. This approach is based on the Kolmogorov-Arnold representation theorem~\cite{KAM1957,Arnold2009}, which asserts that any multivariate continuous function can be represented as a superposition of continuous univariate functions over a compact domain. This theoretical foundation allows KANs to handle complex, multi-parameter systems by breaking them down into simpler, univariate components.

One of the most compelling features of KANs in this scientific context is their capability for symbolic regression. Symbolic regression is a form of regression analysis that seeks mathematical expressions, in symbolic form, that best fit a given dataset. Unlike traditional numeric prediction models that offer a fixed structure and limited insights into underlying patterns, symbolic regression with KANs provides formulaic expressions that reveal the inherent relationships and governing laws within the data. This significantly enhances our understanding and offers a more profound and interpretable model.

The ability to derive symbolic equations directly from data can guide subsequent scientific research by offering clearer insights into the relationships and mechanics at play in nuclear phenomena. Such insights are invaluable for advancing theories, refining experimental designs, and developing new technologies in nuclear science. Hence, KANs do not merely predict but also illuminate the pathway towards a deeper comprehension of the intricate dynamics that define nuclear structures.

In the broader context of artificial intelligence (AI) applications in nuclear physics, there is a growing interest in leveraging machine learning to study the intricate phenomena of nuclear physics and derive formulas directly from data. The liquid drop model, which provides a macroscopic description of the nucleus, has been instrumental in understanding nuclear binding energies. AI techniques, particularly those involving deep learning and symbolic regression, have the potential to learn from vast amounts of nuclear data and derive similar or even more refined mass formulas. These data-driven approaches can capture subtle patterns and correlations that might be overlooked by traditional models, thus offering a complementary perspective to theoretical frameworks. A significant advantage of AI for physics is its ability to uncover hidden relationships within complex datasets. In the current study, we aim to derive mass formulas using AI, harnessing its power to reveal insights that can enhance our understanding of nuclear binding energies.

In this study, we aim to explore the potential of KANs in predicting nuclear binding energies. By leveraging the capabilities of KANs, we seek to bridge the gap between data-driven insights and theoretical understanding, ultimately contributing to the advancement of nuclear science. The paper is organized as follows: In Sec.~\ref{sec:theory}, we provide a brief overview of the Kolmogorov-Arnold network and the data collection and feature selection used in the current study. Sec.~\ref{sec:results} presents the application of KANs to nuclear binding energy. Finally, we conclude with a summary and discussion in Sec.~\ref{sec:sum}.

\section{Formalism}\label{sec:theory}
\subsection{Kolmogorov-Arnold network}
The Kolmogorov-Arnold theorem asserts that any continuous multivariate function on a bounded domain can be decomposed into a finite composition of continuous univariate functions and addition. Specifically, for a continuous function $ f: [0,1]^n \rightarrow \mathbb{R} $:
\begin{equation}
 f(\mathbf{x}) = f(x_1, x_2, \ldots, x_n) = \sum_{q=1}^{2n+1} \Phi_q\left(\sum_{p=1}^n \phi_{q,p}(x_p)\right)   .
 \label{eq:1.1}
\end{equation}
Here, $\mathbf{x} = (x_1, x_2, \ldots, x_n)$ is an $ n $-dimensional vector with each $ x_i $ in $[0,1]$. The functions $\phi_{q,p}: [0,1] \rightarrow \mathbb{R}$ and $\Phi_q: \mathbb{R} \rightarrow \mathbb{R}$ are continuous univariate functions. This decomposition shows that addition is the only truly multivariate operation, as all other multivariate functions can be built using univariate functions and sums. This is useful for identifying learnable and explainable patterns in data. However, practical challenges in machine learning arise due to the potential non-smooth and fractal nature of the functions $\phi_{q,p}$ and $\Phi_q$, complicating decomposition and learning processes.

Recent research by Liu \textit{et al.}~\cite{liu2025kan} has renewed interest in the Kolmogorov-Arnold theorem and its use in neural networks. This has led to the creation of KANs. Liu~\textit{et al.} have improved the original framework by allowing networks to have any number of layers and widths. This overcomes the original theorem's limitations, which restricted it to two layers of nonlinearities and a fixed number of terms $(2n + 1)$. Since most real-world functions are usually smooth and have simple structures, these improvements help create more practical and efficient Kolmogorov-Arnold representations.

\begin{figure}[tb]
\begin{center}
 {\centering \resizebox*{0.8\columnwidth}{!}{\includegraphics{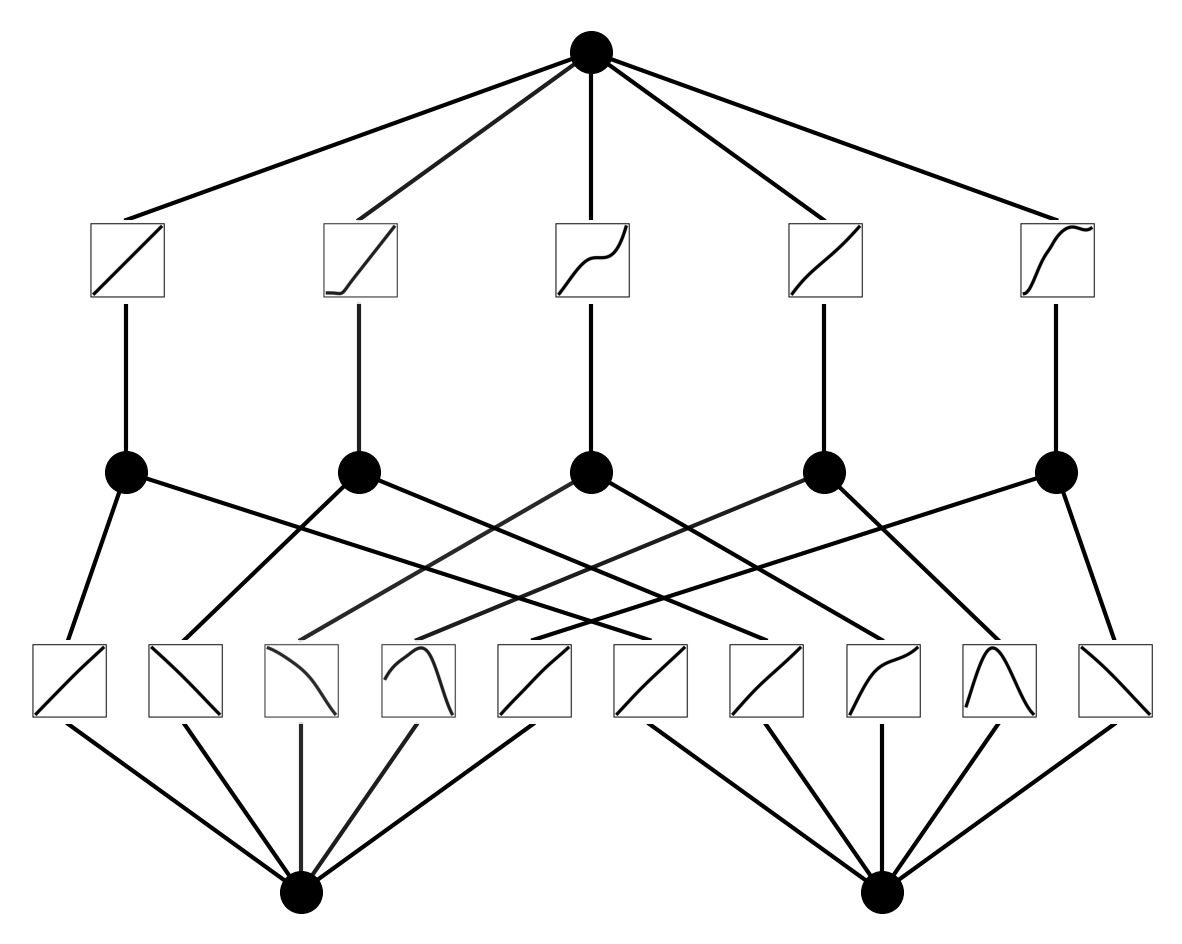}} \par}
\caption{\label{fig:1} Architecture of a shallow KAN, featuring two input features, a hidden layer with five nodes, and a single-node output layer.}
\end{center}
\end{figure}

In Fig.~\ref{fig:1}, a shallow KAN is illustrated. It consists of two input variables, $x_1$ and $x_2$, a hidden layer with five nodes, and a single output node. Each input is connected to the hidden nodes through edges characterized by adaptable functions, which replace the traditional fixed activation functions used in multilayer perceptrons (MLPs).

These adaptable functions are expanded into basis functions, specifically B-spline functions, which dynamically adjust based on the input data. The outputs from the hidden nodes are processed through additional spline-based edges, culminating in a summation at the output node. This structure allows for increasing the network depth, enabling the analysis of more complex physical phenomena.

\subsection{Data collection and feature selection}
In this study, we used mass excess values from the Atomic Mass Evaluation (AME2020)~\cite{Wang_2021}. We focused on nuclei with both the atomic number (Z) and neutron number (N) greater than or equal to 8, covering 3456 nuclei. This selection ensures the model's precision~\footnote{In our study, \textit{accuracy} refers to a qualitative performance characteristic, expressing the closeness of agreement between a measurement result and the value of the measurand, whereas \textit{precision} refers to the closeness of agreement between independent test results obtained under stipulated conditions, as discussed in Ref.~\cite{Menditto2007UnderstandingTM}.} and stability for larger, more complex nuclei.

\begin{figure}[tb]
\begin{center}
 {\centering \resizebox*{0.9\columnwidth}{!}{\includegraphics{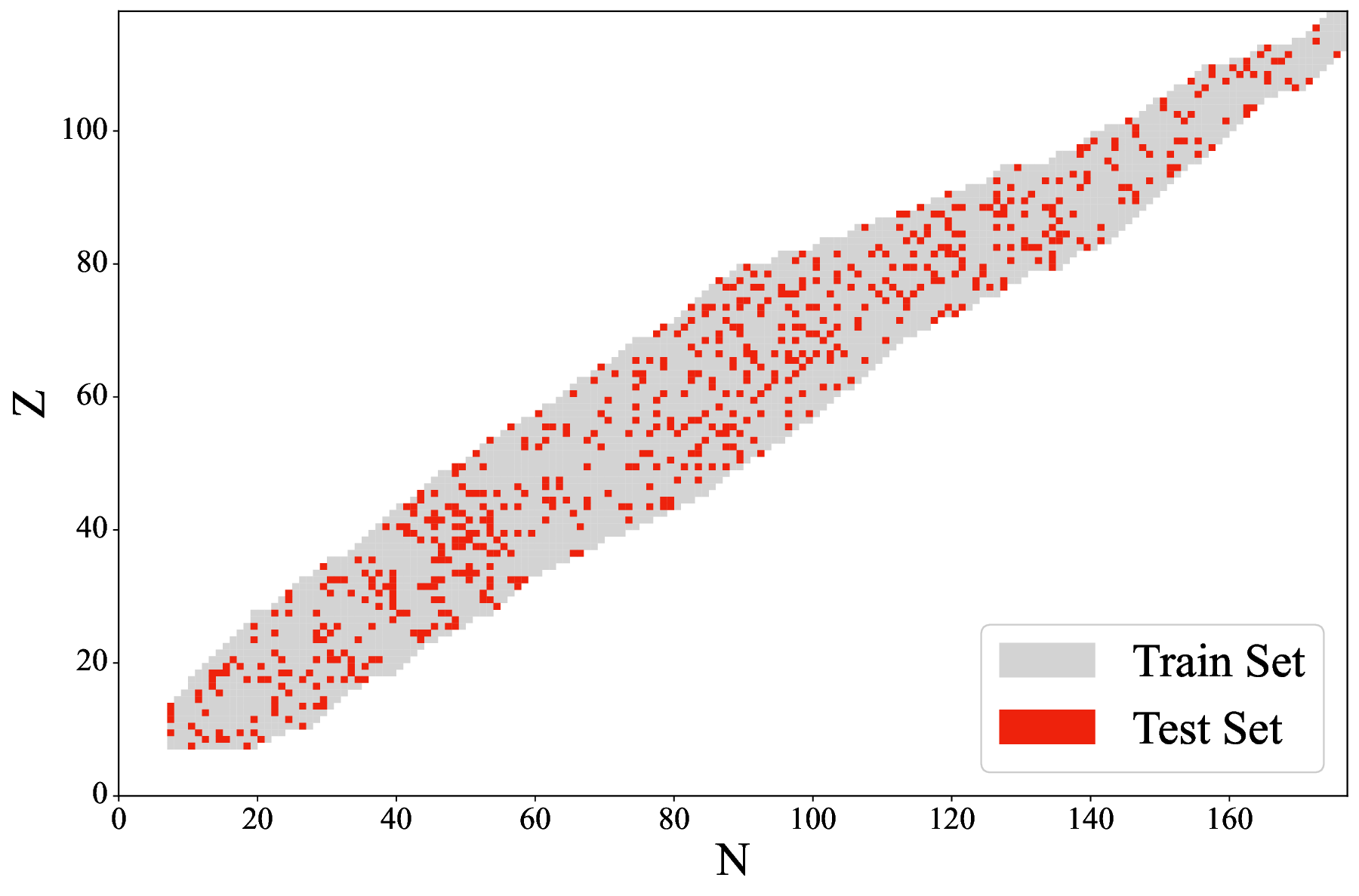}} \par}
\caption{\label{fig:6} The training set (gray circles), test set (red circles) used in the KANs. Both the training and test sets include nuclei from AME2020.}
\end{center}
\end{figure}

The experimental data are randomly divided into two subsets: 2,856 nuclei for training and 600 nuclei for testing. Figure~\ref{fig:6} illustrates the selection of the training and test sets. The nuclei in the training and test sets remain consistent across all calculations.
 
To make the best use of our training set, we applied k-fold cross-validation~\cite{kfold1,kfold2,kfold3}, which ensures that every data point is used for both training and validation, maximizing the use of available data.  Initially, we employed the Fisher-Yates shuffle algorithm~\cite{19531603584} to ensure a thorough and unbiased randomization of the dataset, then split the dataset into $k$ parts (folds). In each iteration, $k-1$ parts are used for training, and the remaining part is used for validation. This process repeats $k$ times, with each part serving as the validation set once.  Averaging the performance across all iterations gives a reliable measure of the model's generalizability. In our calculations, $k$ is set to 3, thereby allocating $1/3$ of the data for testing and the remaining $2/3$ for training in each cycle, with the initial randomization facilitated by the Fisher-Yates shuffle algorithm to ensure fairness and unpredictability in the distribution of data points across folds.

We carefully selected data attributes to ensure the framework's effectiveness and accuracy. These include the atomic number (Z), neutron number (N), and mass number (A). We also included features from previous research~\cite{ml2024,ann2024,WANG2014215}, such as pairing effects, shell closures, and asymmetry, which are crucial for modeling nuclear binding energies.

Pairing effects are captured using $Z_{\text{EO}}$ and $N_{\text{EO}}$, which can only be either 1 or 0. Here, 1 indicates odd $Z/N$ ratios, and 0 indicates even ratios. This distinction helps the model differentiate between various types of nuclei, each possessing unique binding energy characteristics. Additionally, the term $A^{2/3}$ represents the surface effect, reflecting that a nucleus's surface area scales with the two-thirds power of its mass number ($A$). This scaling is crucial for modeling the surface contributions to the nucleus's overall binding energy.

The isospin asymmetry, $(N-Z)$, measures the difference between neutrons and protons. We also consider nuclear magic numbers, with $\mu_Z$ and $\mu_N$ representing the excess numbers of protons and neutrons relative to the nearest closed shell. Key magic numbers include $8, 20, 28, 50, 82, 126, 184$.

We also explore nuclear shells with $Z_{\text{shell}}$ and $N_{\text{shell}}$, reflecting the orbital configurations of the last proton and neutron. These values range from 0 to 5, depending on proton or neutron counts within specific intervals. This helps the model capture the shell structure's influence on binding energy.

We evaluated various combinations of features to identify the most informative attributes for predicting nuclear binding energies. This approach helped us determine the optimal feature set for our KAN framework. Our network structure was shallow, consisting of a single hidden layer with a uniform width of 12 neurons. The detailed features and network structure are presented in Table~\ref{tab:1}. According to the requirements of Eq.~(\ref{eq:1.1}), the number of nodes in the second layer should be at least $2n + 1$. Although this condition can be relaxed based on the work in Ref.~\cite{liu2025kan}, we have chosen to use the maximum value of $n + 1$ as the number of neurons in the hidden layer, which in our case is 12.

To improve model performance and interpretability while reducing overfitting, we used L1 regularization and cross-entropy loss as described in Ref.~\cite{liu2025kan}. The loss function is defined as

\begin{equation}
\begin{split}
\text{Loss} =& \frac{(\mathrm{BE}_{\text{pred}} - \mathrm{BE}_{\text{exp}})^2}{N_p} \\
&+ \lambda \left( \mu_1 \sum_{l=0}^{L-1} \left|\boldsymbol{\Phi}_l\right|_1 + \mu_2 \sum_{l=0}^{L-1} S\left(\boldsymbol{\Phi}_l\right) \right),
\end{split}
\end{equation}

where $\left|\boldsymbol{\Phi}_l\right|_1$ is the L1 norm of the KAN layer, defined as the sum of the L1 norms of all activation functions $\phi_{i,j}$, including $n_{\text{out}}$ outputs and $n_{\text{in}}$ inputs. The L1 norm of an activation function, averaged over $N_p$ inputs, is given by

\begin{equation}
|\phi|_1 \equiv \frac{1}{N_p} \sum_{s=1}^{N_p} \left| \phi\left(x^{(s)}\right) \right|.
\end{equation}

Thus, the L1 norm of a KAN layer is

\begin{equation}
|\boldsymbol{\Phi}|_1 \equiv \sum_{i=1}^{n_{\text{in}}} \sum_{j=1}^{n_{\text{out}}} \left|\phi_{i,j}\right|_1.
\end{equation}

The entropy of a KAN layer is defined as

\begin{equation}
S(\boldsymbol{\Phi}) \equiv -\sum_{i=1}^{n_{\text{in}}} \sum_{j=1}^{n_{\text{out}}} \frac{\left|\phi_{i,j}\right|_1}{|\boldsymbol{\Phi}|_1} \log \left( \frac{\left|\phi_{i,j}\right|_1}{|\boldsymbol{\Phi}|_1} \right).
\end{equation}

L1 regularization helps make the model simpler by adding a penalty based on the absolute values of the model’s parameters. This leads to models with fewer nonzero parameters, making them easier to interpret and less likely to overfit. Cross-entropy loss ensures the model’s predictions are well calibrated by measuring the difference between predicted and actual probability distributions.

Combining these techniques, we aim to create a robust and understandable model for predicting nuclear binding energies. We used k-fold cross-validation~\cite{kfold1,kfold2,kfold3} to test different parts of the data, selected features carefully, and applied advanced regularization techniques to ensure our KAN framework is precise and generalizable. By exploring different features and network structures, we gain insights into what affects model performance, helping us develop better neural network architectures for nuclear physics.

\begin{table}[ht]
	\centering
	\caption{Feature space and KANs' structure}
	\label{tab:1}  
	\begin{tabular}{cccccccc}
		\noalign{\smallskip}\hline\noalign{\smallskip}
        Model & & Feature & & Structure\\
        \noalign{\smallskip}\hline\noalign{\smallskip}
		KAN-2 & & N, Z  & &[2,12,1]\\
        KAN-4 & & N, Z, A, N-Z & &[4,12,1]\\
		KAN-9 & & N, Z, A, N-Z, $A^{2/3}$, $Z_{\text{EO}}$, $N_{\text{EO}}$, & &[9,12,1]\\
              & & $\mu_Z$, $\mu_N$   \\
		KAN-11 & & N, Z, A, N-Z, $A^{2/3}$, $Z_{\text{EO}}$, $N_{\text{EO}}$,& &[11,12,1] \\
              & & $\mu_Z$, $\mu_N$, $Z_{\text{shell}}$, $N_{\text{shell}}$\\
		\noalign{\smallskip}\hline
	\end{tabular}
\end{table}

\section{Results}\label{sec:results}
\begin{figure*}[tb]
\begin{center} 
 {\centering \resizebox*{2.0\columnwidth}{!}{\includegraphics{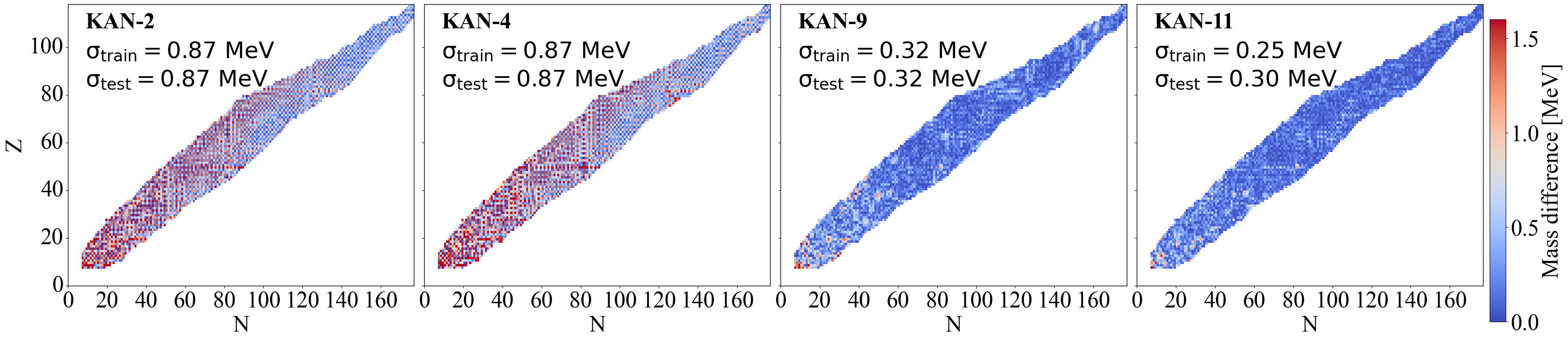}} \par}
\caption{\label{fig:2} The absolute value of binding energy differences between KAN predictions using different features (see Table~\ref{tab:1}) and AME2020~\cite{Wang_2021}. The RMSEs for all the data are also provided.}
\end{center}
\end{figure*}
\begin{figure*}[tb]
\begin{center} 
 {\centering \resizebox*{2.0\columnwidth}{!}{\includegraphics{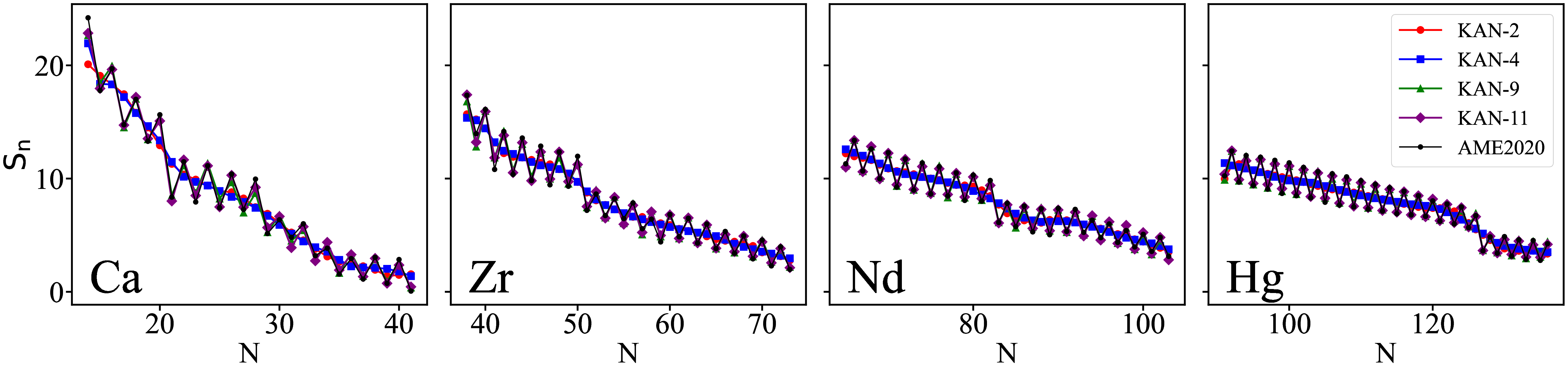}} \par}
\caption{\label{fig:3} Experimental single neutron separation energies for $\mathrm{Ca}$, $\mathrm{Zr}$, $\mathrm{Nd}$, and $\mathrm{Hg}$ isotopic chains in comparison with the KAN predictions.}
\end{center}
\end{figure*}
Figure~\ref{fig:2} shows the absolute differences in binding energy between predictions from various KAN models and the experimental data from AME2020~\cite{Wang_2021}. The feature sets used in these models are listed in Table~\ref{tab:1}. Notably, the KAN-2 model, which uses only basic properties such as neutron number $N$ and proton number $Z$, achieves a root mean square error (RMSE) of just 0.87 MeV. This performance surpasses that of many microscopic models~\cite{ml2024}, highlighting the effectiveness of the KAN approach even with minimal input features. When additional features are incorporated, the KAN-11 model reduces the RMSE even further to an impressive 0.26 MeV for the entire data set. Furthermore, the nearly identical RMSEs for both the training and test sets indicate the robustness achieved through the k-folding method and KANs.

As more features, such as isospin asymmetry and mass number, are included, the RMSE decreases, as illustrated in Fig.~\ref{fig:2}, progressing from KAN-2 to KAN-4. However, both models (KAN-2 and KAN-4) show a significant increase in deviation error as nuclei approach closed shells. This error distribution underscores the influence of shell effects on nuclear binding energy. By incorporating features related to shell structure, models KAN-9 and KAN-11 achieve lower RMSEs. Figure~\ref{fig:2} demonstrates that these models perform better for medium and heavy nuclei compared to light nuclei. Despite efforts to enhance predictions by expanding the feature space, the challenge of modeling light nuclei persists.

While we acknowledge that increasing the total number of parameters can introduce complexity and potential issues, it's important to analyze this impact carefully. In our models, we selected 30 basis functions to represent each activation function. Although using more basis functions can better approximate the ``true activation function,'' it also increases the model's parameter count and computational demands. The required number of basis functions largely depends on the properties of the selected basis functions and the behavior of the activation functions they aim to approximate. Therefore, selecting appropriate basis functions tailored to the specific problem and dataset can significantly reduce the number of parameters without compromising performance.

Previous research has explored various alternative basis functions beyond traditional B-spline curves. Researchers have experimented with sine functions, trainable adaptive fractional-orthogonal Jacobi functions, wavelets, polynomials, and rational functions based on Padé approximations and rational Jacobi functions~\cite{aghaei2024fkan,aghaei2024rkanrationalkolmogorovarnoldnetworks,warin2024p1kaneffectivekolmogorovarnold,10.3389/frai.2024.1462952}. For instance, the development of rational Kolmogorov-Arnold networks (rKANs) utilizes rational functions as basis functions, offering improved approximation capabilities while more effectively managing the parameter scale~\cite{aghaei2024rkanrationalkolmogorovarnoldnetworks}. These alternative basis functions enhance the network's ability to model complex functions while optimizing computational efficiency. Therefore, in KANs, the specific choice of basis functions is not as critical; rather, the number of activation functions plays a more significant role. Therefore we set the same width for the hidden layer.

For all that, we can still compare the parameter scales of different KANs. To assess the parameter scale, we note that KAN-2 roughly contains 1080 free parameters; KAN-4 has about 1800; KAN-9 has 3000; and KAN-11 has 4320 free parameters. As mentioned earlier, for KAN-2 and KAN-4, we deliberately increased the width of their hidden layers to match the network scale of the more complex models. In practical applications, similar RMSE can be achieved with a narrower width. For example, for KAN-2, using architectures like [2,5,1] or [2,12,1], we achieved comparable RMSE on the dataset, and the parameter count of [2,5,1] is only 450—less than half of the original. Additionally, by applying L1 regularization, we can prune unimportant branches to obtain a smaller model, as demonstrated in Sec.~\uppercase\expandafter{\romannumeral3}~C.

We also compare our results with previous work. A recent application of mixture density networks (MDNs), a type of machine learning algorithm, resulted in errors of about 0.5--0.6 MeV compared to AME2016 when combined with physics-based features~\cite{PhysRevC.106.014305}. Gaussian process regression (GPR) achieved errors ranging from 0.14 to 0.96 MeV for the training set and 0.26 to 1.08 MeV for the test set compared to AME2020, using different physical features as reported in Ref.~\cite{ml2024}. Support vector regression (SVR) reached errors from 0.39 to 2.55~MeV for the test set and from 0.23 to 2.40~MeV for the training set compared to AME2020, as shown in Ref.~\cite{ml2024}. 

Meanwhile, Ref.~\cite{PhysRevC.106.L021303} shows that Bayesian neural networks (BNNs) can achieve a RMSE of 84~keV. They used even-even nuclei as the foundation of their dataset and determined the binding energies of surrounding nuclei by simultaneously predicting and combining the separation energies of neighboring nuclei. Compared to these models, KAN achieves similar RMSE, except for the BNN, which exhibits superior performance. This result has already surpassed the performance of most model-driven methods. For example, the corresponding RMSEs are 0.285, 0.559, and 0.576~MeV for the WS4~\cite{WANG2014215}, FRDM2012~\cite{MOLLER20161}, and HFB-31~\cite{PhysRevC.93.034337} models, respectively. Given that KAN is a novel technology, there is reason to believe that ongoing developments will further enhance its capabilities. As the technology matures, we can expect predictions to become even more accurate and precise, leading to broader applications in nuclear physics and machine learning.

Fairly comparing training times among different models is challenging due to variations in implementation, computational resources, and optimization strategies. However, as noted in Ref.~\cite{liu2025kan}, while KANs offer the advantage of reducing network size compared to traditional MLPs, MLPs tend to be more efficient in training networks of the same scale. However, KANs possess distinct properties, such as the ability to initiate training with a fewer number of basis functions for approximating activation functions, gradually refining them as needed. This capability allows us to start with a simpler model to capture the fundamental shape of activation functions and incrementally increase the basis functions to enhance detail. This approach simplifies the training process and may provide benefits in terms of interpretability and computational efficiency. Furthermore, recent studies~\cite{aghaei2024fkan,aghaei2024rkanrationalkolmogorovarnoldnetworks,warin2024p1kaneffectivekolmogorovarnold,10.3389/frai.2024.1462952} have demonstrated that various KAN variants are becoming increasingly efficient, potentially paving the way for more effective implementations in diverse applications.

Figure~\ref{fig:3} compares the separation energies for isotopic chains of four different elements against the values in AME2020. The absence of pairing terms in KAN-2 and KAN-4 leads to much larger prediction RMSEs compared to KAN-9 and KAN-11, which include these terms. This difference is especially noticeable in the color variations representing the absolute differences in binding energy between the predictions from various KAN models and the experimental data from AME2020, as shown in Fig.~\ref{fig:2}. These variations are present in KAN-2 and KAN-4 but not in KAN-9 and KAN-11, highlighting the crucial role of pairing effects in detailing nuclear structure. The single-neutron separation energy curves in Fig.~\ref{fig:3} further demonstrate the impact of pairing. The difficulties faced by KAN-2 and KAN-4 models with abrupt changes suggest that adding features like $N_{EO}$ and $Z_{EO}$ could better model the pairing behavior in nuclear binding energies, leading to improved performance in KAN-9 and KAN-11.

However, small differences remain in the calcium ($\mathrm{Ca}$) isotopic chain, as shown in Fig.~\ref{fig:3}. The model shows larger RMSEs for light nuclei compared to medium and heavy nuclei. Light nuclei are more affected by interactions among a few particles, while heavier nuclei are influenced by interactions among many particles. This difference makes it challenging to create a model that works well for all types of nuclei. It suggests that we need a more advanced KAN model to better understand the complex behavior of nuclear systems across different mass regions.

\subsection{Extrapolation}

\begin{figure*}[tb]
\begin{center} 
 {\centering \resizebox*{2.0\columnwidth}{!}{\includegraphics{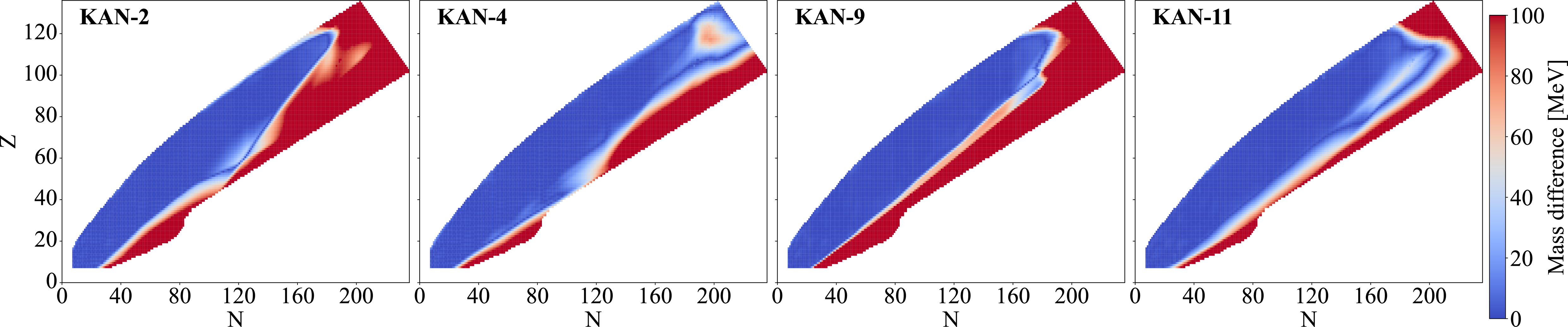}} \par}
\caption{\label{fig:7} The absolute value of mass excess differences between KAN predictions using different features (see Table~\ref{tab:1}) and FRDM12's prediction~\cite{MOLLER20161}. }
\end{center}
\end{figure*}
\begin{figure*}[tb]
\begin{center} 
 {\centering \resizebox*{2.0\columnwidth}{!}{\includegraphics{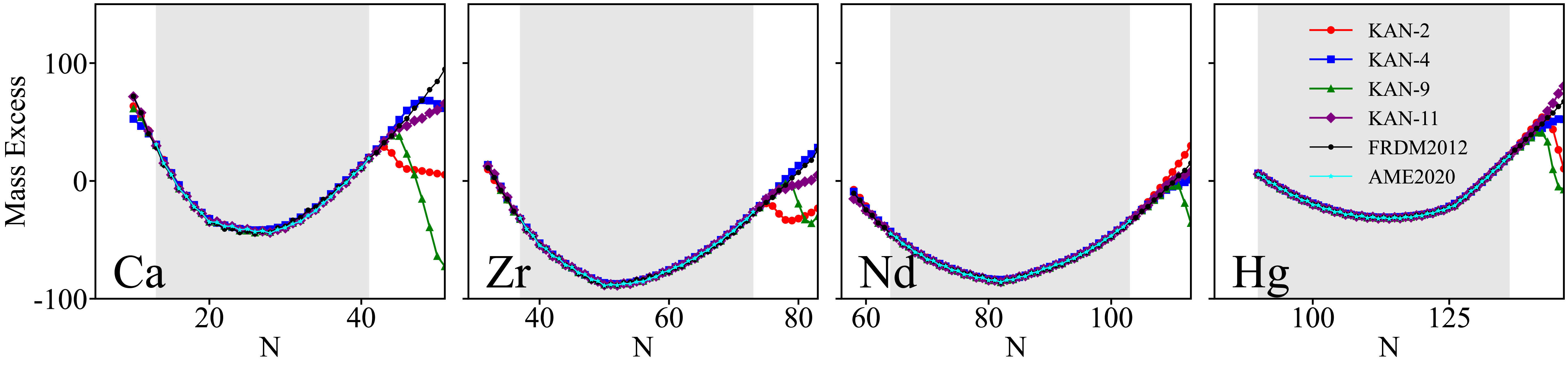}} \par}
\caption{\label{fig:8} Mass excess for $\mathrm{Ca}$, $\mathrm{Zr}$, $\mathrm{Nd}$, and $\mathrm{Hg}$ isotopic chains compared with KANs predictions, FRDM12 (black points)~\cite{MOLLER20161}, and experimental data from AME2020 (cyan stars)~\cite{Wang_2021}.}
\end{center}
\end{figure*}

Extrapolating beyond the training dataset presents a significant challenge for machine learning and deep learning models. To evaluate the extrapolation capability of our KANs, we initially focused on nuclei at the edges of the AME2020 dataset included in the test set—specifically, boundary isotopes and isotones such as $\mathrm{^{92}_{59}As}$, $\mathrm{^{91}_{58}As}$, $\mathrm{^{90}_{57}As}$, and $\mathrm{^{215}_{135}Hg}$, $\mathrm{^{216}_{135}Tl}$, $\mathrm{^{217}_{135}Pb}$. We observed that KANs could predict the binding energies of these nuclei with reasonable RMSEs. However, since these nuclei are very close to the data in the training set, a more thorough assessment of the KANs' ability to extrapolate was deemed necessary.

To accomplish this, we compared the predictions of our KANs with those from the finite-range droplet model (FRDM2012)~\cite{MOLLER20161}. Figure~\ref{fig:7} showcases the mass excess predictions made by KANs with various feature sets alongside the results from FRDM2012. This dataset encompasses 9138 nuclei, ranging from $\mathrm{^{16}O}$ up to a mass number $A = 339$, significantly extending beyond the scope of AME2020.

Our observations revealed that the KAN-4 and KAN-11 models demonstrated good agreement with FRDM2012 across most regions, with many differences between these models and FRDM2012 being within 50 MeV. Considering we are dealing with nuclei far from the known mass boundaries, such discrepancies between different models are deemed acceptable. The KAN-2 model, however, exhibited limited extrapolation capability, which is understandable given its minimal feature set. Interestingly, the KAN-9 model showed signs of overfitting to a certain extent.

Furthermore, in regions significantly distant from the training data, our KAN models exhibit larger deviations from the results predicted by the physically motivated FRDM2012 model. This discrepancy can be attributed to two main factors. First, the divergence among different models tends to increase when extrapolating far beyond the known data regions. This phenomenon is also observed in Refs.~\cite{ml2024,PhysRevC.98.034318}, where differences among various density functional models and machine learning models are noted. Second, the use of B-splines as basis functions for expanding the activation function means the model's behavior beyond the training set heavily relies on the edge behavior of the B-spline curves. Essentially, the extrapolation is influenced more by the polynomial behavior at the boundaries rather than by the intended activation functions.

A potential solution to improve extrapolation capabilities could involve employing symbolic regression to replace the polynomial expansion form. Functions derived from symbolic regression may inherently possess better extrapolation capabilities due to their flexibility and the ability to capture underlying patterns more effectively.

On the other hand, as depicted in Fig.~\ref{fig:8}, when predicting isotopic chains extending ten neutrons beyond the existing dataset, the KANs' predictions closely align with those from FRDM2012. This suggests that the KAN networks have reliable extrapolation abilities in regions not too distant from the training data, demonstrating stable and trustworthy performance. It is also evident that KAN performs better on heavy nuclei than on light nuclei, consistent with the observations made in Figs.~\ref{fig:2} and \ref{fig:3}.

\subsection{Feature analysis}
\begin{figure}[ht]
    \centering
    \includegraphics[width=1.0\linewidth]{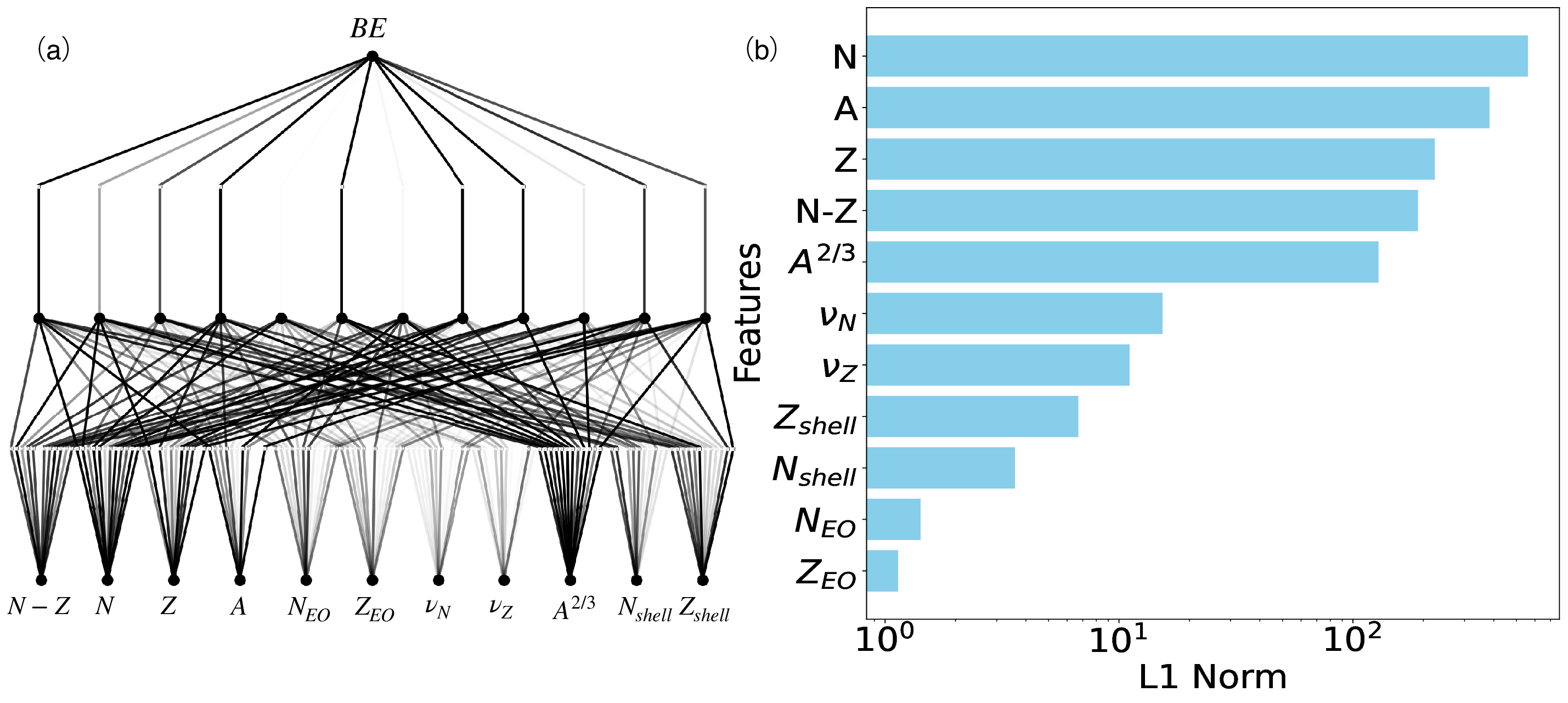}
    \caption{Visualization of the trained network structure in the panel (a). The transparency of the lines connecting nodes is proportional to $\tanh(\beta A_{l,i,j})$, where $A_{l,i,j}$ is the mean activation, enhancing the visibility of connections. In panel (b), the $ L1 $ norm for each connection is shown on a logarithmic scale, highlighting the significance of each link.}
    \label{fig:4}
\end{figure}
In Fig.~\ref{fig:4} we show the structure of the trained KAN. To clarify node connections, we adjust the transparency of lines based on the mean value of the activation function $\phi_{l,i,j}$ for each connection, represented by $\tanh(\beta A_{l,i,j})$, where $A_{l,i,j}$ is the mean activation. This technique highlights each variable's influence on the final output, aiding in understanding the network's behavior. Additionally, Fig.~\ref{fig:4}(b) illustrates the $L1$ norm for each connection, providing insights into the importance of each link and identifying significant contributors to the model's performance. This dual visualization of mean activation and $L1$ norm offers a comprehensive view of the network's workings and the relative importance of different features and connections.

Our analysis indicates that the neutron number $N$ exerts the largest influence on the model's predictions, as evidenced by its highest $L1$ norm among the variables. The nucleon number $A$ and proton number $Z$ also significantly impact the model, with their $L1$ norms being large and comparable to that of $N$. This suggests that $N$, $A$, and $Z$ are the primary factors influencing nuclear properties in our model.

The asymmetry term $N-Z$ and the surface term $A^{2/3}$ exhibit moderate effects, highlighting their relevance but lesser dominance compared to $A$, $N$, and $Z$. These terms are crucial for capturing the nuances of nuclear structure, though their impact is not as pronounced as the primary variables.

Variables associated with nuclear shell structure and the pairing term demonstrate minimal direct impact on the overall results. Despite their subtle influence, these factors are essential for accurate predictions of nuclear binding energies, especially in astrophysics and nucleosynthesis, where high accuracy and precision are paramount. Figures \ref{fig:2} and \ref{fig:3} show that incorporating these terms, although they contribute less directly, is vital for modeling.

These findings align with our calculations and are supported by results from other studies \cite{ann2024,ml2024,WANG2014215,PhysRevC.106.014305,PhysRevC.106.L021301}, reinforcing the validity of our approach and the robustness of our model. This analysis underlines the primary drivers of nuclear binding energy and emphasizes the importance of considering all relevant factors, even those with less apparent impact, to accurately capture the complexity of nuclear interactions.

It is important to note, however, that this analysis can only provide approximate relationships between the variables. Due to the inherent randomness in the training process—such as the random initialization of model weights and the stochastic nature of optimization algorithms—the exact values of the $L1$ norms may vary across different training sessions. This means that while we can identify general patterns and trends, establishing absolute relationships between the variables based on a single instance of training is not possible.

\subsection{Symbolic regression}
One of the most intriguing aspects of Kolmogorov-Arnold networks (KANs) is their ability to directly derive symbolic expressions from data, which enhances both the interpretability and transparency of the predictive model. Unlike traditional neural networks that employ fixed activation functions—potentially obscuring underlying patterns—KANs learn activation functions tailored to the data. These can initially be represented using B-splines, which are smooth, piecewise-defined functions.

To extract analytical formulas for nuclear binding energy from these learned activation functions, we establish a function library composed of elementary functions and basic functional forms, such as polynomials, exponentials, and logarithms. By utilizing the least-squares method, we adjust the coefficients of these functions to best match the learned activation functions, effectively performing symbolic regression. Afterward, by combining each activation function through Eq.~(\ref{eq:1.1}), we obtain the final expression, with specific details available in Ref.~\cite{liu2025kan}.

To simplify the model and reduce complexity introduced by periodic functions (like sine functions that may be influenced by pairing terms), we focus on even-even nuclei. This selection minimizes the influence of pairing effects on the activation functions. Additionally, we limit our study to medium to heavy nuclei, with proton and neutron numbers $Z \geq 30$ and $N \geq 30$, as these nuclei exhibit more uniform behavior conducive to symbolic modeling.

We initially trained a KAN with a straightforward architecture: two input neurons, one hidden layer with five neurons, and one output neuron. To enhance network sparsity and maintain performance while increasing the regularization coefficient, we pruned connections with minimal impact on the output. This pruning simplified the network to one with two input neurons, one hidden layer containing two neurons, and one output neuron. Remarkably, the final model achieved a RMSE of 0.66 MeV. With this streamlined network, we proceeded to perform symbolic regression on the learned activation functions.

It's important to acknowledge that approximating each activation function with elementary mathematical functions inevitably introduces some error. These approximation errors can accumulate and propagate through the network, potentially leading to larger discrepancies in the final predictions compared to the original KAN model. Consequently, the symbolic regression results may exhibit greater errors relative to the initial model. Nevertheless, the derived analytical expressions provide valuable insights into the underlying physical laws governing nuclear binding energies.

\begin{figure}[h]
    \centering
    \includegraphics[width=1.0\linewidth]{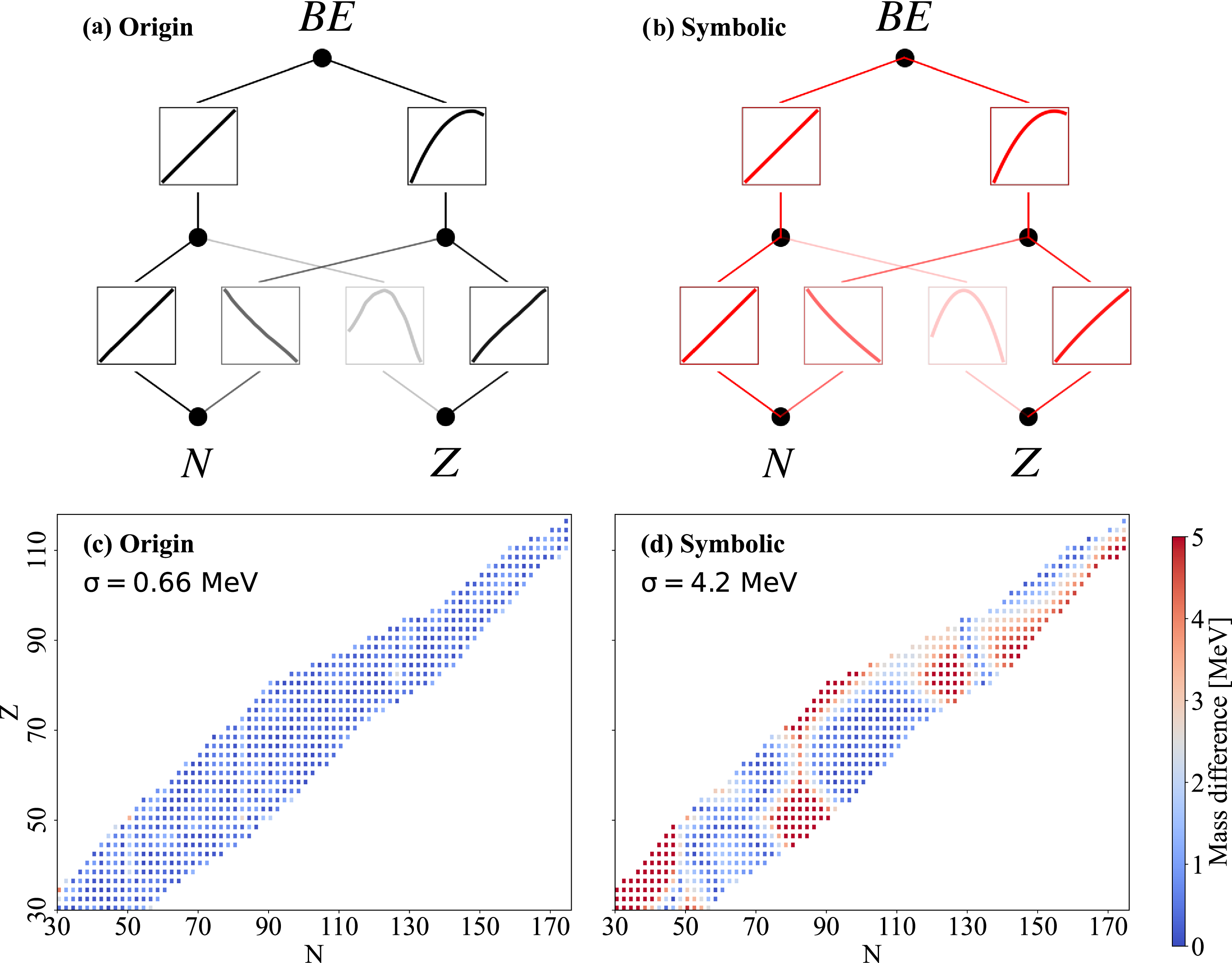}
    \caption{Panel (a) shows the original trained network structure, while panel (b) shows the symbolic network structure. Panels (c) and (d) illustrate the absolute value of binding energy differences between predictions and data from AME2020~\cite{Wang_2021}.}
    \label{fig:5}
\end{figure}

As illustrated in Fig.~\ref{fig:5}, transitioning from the original network's activation functions [the black line in Fig.~\ref{fig:5}(a)] to the symbolic network's activation functions [the red line in Fig.~\ref{fig:5}(b)] results in some loss of detail. These lost details propagate through the network, leading to varying degrees of error in the predictions. A comparison of the differences between the original and symbolic models against experimental data [Figs.~\ref{fig:5}(c) and \ref{fig:5}(d)] reveals underlying patterns: near magic numbers 50, 82, and 126, the errors in the symbolic results are notably larger than those in the original results. This suggests that the nuances not fully captured by the symbolic approximations in the activation functions reflect structural information about the nuclei—information that is crucial near these magic numbers due to shell closures and associated nuclear structure effects.

Despite not including variables explicitly related to shell structure in our input features, the KAN demonstrated a robust capability to identify and learn these features directly from the data. This ability underscores the strength of KANs in capturing complex physical phenomena and validates the effectiveness of our symbolic regression approach in revealing and confirming underlying physical laws, even when working with simplified models and limited input variables.

Below is the expression obtained through symbolic regression. After simplification, it can be presented in the following form, and the parameters are listed in Table~\ref{tab:2}:
\begin{equation}
\begin{aligned}
    BE(N,Z)=&A_{vol} A  - A_{sym} (N-Z) - A_{coul} Z^2\\
    &-[A_N(N+n)^{2/3}-A_Z(Z+z)^{2/3}+C_2]^2-C_1.
\label{eq:2.1}
\end{aligned}
\end{equation}

\begin{table}[h]
	\centering
	\caption{Parameter and its value for Eq.~(\ref{eq:2.1}).}
	\label{tab:2}  
	\begin{tabular}{clccccccccccc}
		\noalign{\smallskip}\hline\noalign{\smallskip}
        Parameter &Value & Parameter &Value\\
        \noalign{\smallskip}\hline\noalign{\smallskip}
		$A_{vol}~(\mathrm{MeV})$ &9.87917 & $A_N~(\mathrm{MeV^{1/2}})$ & 0.53670 \\
        $A_{sym}~(\mathrm{MeV})$ &-1.56157 & $A_Z~(\mathrm{MeV^{1/2}})$& 2.93091\\
		$A_{coul}~(\mathrm{MeV})$ &0.060169 & $n$& -1.06419\\
        $C_1~(\mathrm{MeV})$ &13.08641 & $z$& 6.26334 \\
		$C_2~(\mathrm{MeV^{1/2}})$ &11.76810 & & \\

		\noalign{\smallskip}\hline
	\end{tabular}
\end{table}

This expression for the binding energy achieves an RMSE of 4.2 MeV for these even-even nuclei. Given that the binding energies for nuclei in this range lie between 500 and 2100 MeV, this suggests that the formula retains less than
1\% error.

When compared to the classical liquid drop model and the Bethe-Weizsäcker formula, the initial three terms in the equation align with the volume, symmetry energy, and Coulomb terms, respectively. The fourth term involves the square of the difference between the cube roots of the neutron and proton numbers, reflecting the surface energy effects of the nucleus. Here, $A_N$ and $A_Z$ are the surface energy coefficients for neutrons and protons, respectively, and $n$ and $z$ are adjustments to the neutron and proton numbers to capture more subtle nuclear structural effects. $C_2$, as a translation constant, adjusts the overall level of the surface energy term. The final constant, $C_1$, shifts the entire energy expression to optimize the model fit.

\begin{figure*}[tb]
\begin{center} 
 {\centering \resizebox*{2.0\columnwidth}{!}{\includegraphics{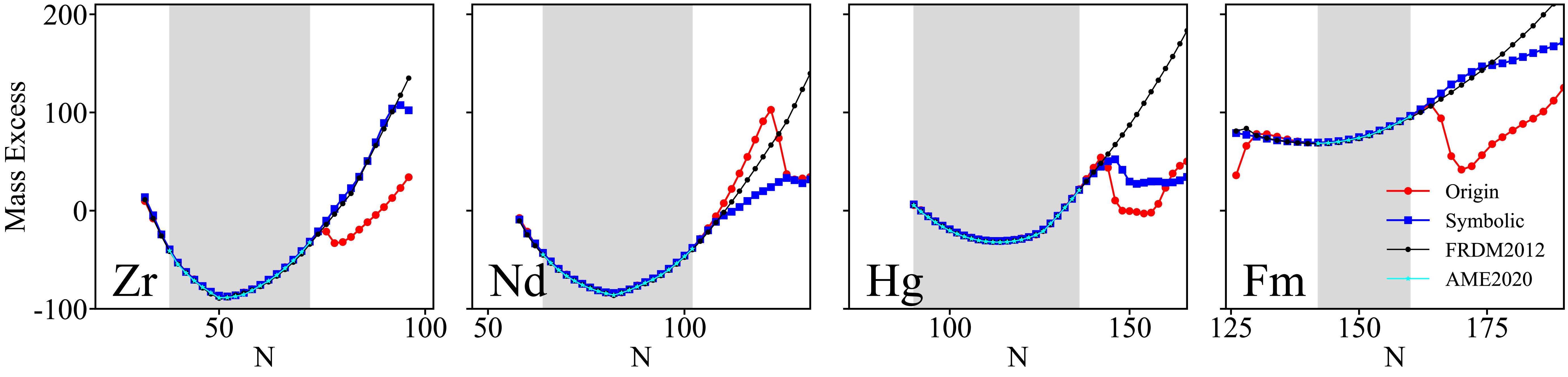}} \par}
\caption{\label{fig:9} Mass excess for $\mathrm{Zr}$, $\mathrm{Nd}$, $\mathrm{Hg}$, and $\mathrm{Fm}$ isotopic chains compared with KANs predictions, the results from symbolic regression, FRDM12 (black points)~\cite{MOLLER20161}, and experimental data from AME2020 (cyan stars)~\cite{Wang_2021}.}
\end{center}
\end{figure*}

It is also worth noting that an underlying relationship exists between $N$ and $Z$ due to the stability and other properties of nuclei. Classical distributions of nuclei often adhere to certain $N/Z$ ratios, as suggested by the Bethe-Weizsäcker formula. This implies that $N$ and $Z$ are not entirely independent variables and can, to some extent, be interchanged. Therefore, the derived analytical expression may not be unique.

As observed in Fig.~\ref{fig:9} and consistent with our previous analysis, the application of symbolic regression enhances the stability of extrapolation. For the isotopic chains of Zr, Nd, Hg, and Fm, the predictions obtained through symbolic regression (depicted as blue squares) exhibit greater stability compared to those from the original KAN approach (represented by red dots). This indicates that symbolic regression provides corrective effects of varying magnitudes at the boundaries of the data, thereby improving the model's predictive performance in these regions.

However, it is important to acknowledge that symbolic regression cannot entirely eliminate the issues associated with extrapolating beyond the range of the available data. Such extrapolation can lead to excessively weak constraints on the predicted results, resulting in potential model failures. Therefore, while symbolic regression contributes to improved extrapolation stability, it does not fully overcome the inherent limitations of extrapolation in machine learning models. Care must still be taken when making predictions outside the scope of the training data to ensure the reliability of the results.

\section{\label{sec:sum} Summary and conclusions}
In this study, we have demonstrated the effectiveness of KANs in predicting nuclear binding energies. By leveraging the capabilities of KANs to decompose complex multivariate functions into univariate components, we have shown that even with minimal input features, KANs can achieve significantly lower RMSE, surpassing many traditional physics-based models. The incorporation of additional features related to isospin asymmetry, shell structure, and pairing effects further reduces the RMSE of the predictions, emphasizing the importance of these factors in modeling nuclear binding energies.

Our results show that KANs can effectively capture the underlying physical relationships in nuclear data, achieving a RMSE as low as 0.26 MeV for the entire dataset when using an expanded set of features. The KAN models exhibit robust performance across different regions of the nuclear chart, particularly for medium and heavy nuclei. However, challenges remain in modeling light nuclei, suggesting the need for more advanced models or additional features to capture the complexities associated with few-body dynamics.

The symbolic regression analysis conducted in this work yielded an analytical expression for nuclear binding energy that aligns well with established models like the Bethe-Weizsäcker formula. This expression incorporates key terms such as volume energy, symmetry energy, Coulomb energy, and surface energy, providing direct insight into the functional dependencies within the nuclear data. While the symbolic model exhibited marginally higher RMSE compared to the original KAN model, it offers the advantage of improved interpretability and potential for deeper theoretical understanding.

We also explored the extrapolation capabilities of KANs by comparing their predictions with those from the finite-range droplet model (FRDM2012)~\cite{MOLLER20161} for nuclei beyond the known mass boundaries. The KAN models demonstrated reasonable agreement with FRDM2012 for nuclei not too distant from the training data, indicating stable and trustworthy performance in extrapolation within similar regions. However, larger deviations were observed when predicting nuclei far beyond the training set, highlighting inherent limitations in extrapolation and the need for caution when predicting properties of nuclei in uncharted regions.

Overall, our study highlights the potential of KANs in nuclear physics applications, combining data-driven approaches with theoretical insights to enhance predictive capabilities. The ability of KANs to balance the error with model simplicity and interpretability makes them a promising tool for exploring complex systems where traditional methods may face challenges.

Future work will focus on extending the application of KANs to better model light nuclei and refine the network architectures for improved performance. Additionally, further exploration of symbolic regression techniques within the KAN framework could lead to new theoretical developments and more accurate models in nuclear physics and other fields involving complex multivariate relationships. 

An important direction for future research is the incorporation of uncertainty quantification, as demonstrated in recent advancements with KANs~\cite{giroux2024uncertaintyquantificationbayesianhigher}. By employing a Bayesian approach and enhancing computational efficiency with methods like those proposed for ReLU-KANs, both epistemic and aleatoric uncertainties can be effectively quantified~\cite{giroux2024uncertaintyquantificationbayesianhigher}. This not only broadens the applicability of KANs to nuclear physics by ensuring accurate identification of functional dependencies in stochastic environments, but also enhances the reliability and robustness of predictions. Implementing these methods could pave the way for innovative applications in nuclear physics, providing insights into the model's uncertainty characteristics and revealing areas for improvement.

\section*{Acknowledgments}
This work has been supported by National Natural Science Foundation of China (Grants No. 12475132, No.12105204, and No.12035011), by the National Key R\&D Program of China (Contracts No. 2023YFA1606503), and by the Fundamental Research Funds for the Central Universities.

\bibliography{reference}

@article{Wang_2021,
doi = {10.1088/1674-1137/abddaf},
url = {https://dx.doi.org/10.1088/1674-1137/abddaf},
year = {2021},
month = {mar},
publisher = {Chinese Physical Society and the Institute of High Energy Physics of the Chinese Academy of Sciences and the Institute of Modern Physics of the Chinese Academy of Sciences and IOP Publishing Ltd},
volume = {45},
number = {3},
pages = {030003},
author = {Meng Wang and W.J. Huang and F.G. Kondev and G. Audi and S. Naimi},
title = {The AME 2020 atomic mass evaluation (II). Tables, graphs and references*},
journal = {Chin. Phys. C}
}

@article{ml2024,
  title = {Nuclear mass predictions using machine learning models},
  author = {Y\"uksel, Esra and Soydaner, Derya and Bahtiyar, H\"useyin},
  journal = {Phys. Rev. C},
  volume = {109},
  issue = {6},
  pages = {064322},
  numpages = {11},
  year = {2024},
  month = {Jun},
  publisher = {American Physical Society},
  doi = {10.1103/PhysRevC.109.064322},
  url = {https://link.aps.org/doi/10.1103/PhysRevC.109.064322}
}

@article{ann2024,
  title = {Nuclear binding energies in artificial neural networks},
  author = {Zeng, Lin-Xing and Yin, Yu-Ying and Dong, Xiao-Xu and Geng, Li-Sheng},
  journal = {Phys. Rev. C},
  volume = {109},
  issue = {3},
  pages = {034318},
  numpages = {9},
  year = {2024},
  month = {Mar},
  publisher = {American Physical Society},
  doi = {10.1103/PhysRevC.109.034318},
  url = {https://link.aps.org/doi/10.1103/PhysRevC.109.034318}
}

@article{WANG2014215,
title = {Surface diffuseness correction in global mass formula},
journal = {Phys. Lett. B},
volume = {734},
pages = {215-219},
year = {2014},
issn = {0370-2693},
doi = {https://doi.org/10.1016/j.physletb.2014.05.049},
url = {https://www.sciencedirect.com/science/article/pii/S037026931400358X},
author = {Ning Wang and Min Liu and Xizhen Wu and Jie Meng}
}

@article{PhysRevC.106.L021301,
  title = {Physically interpretable machine learning for nuclear masses},
  author = {Mumpower, M. R. and Sprouse, T. M. and Lovell, A. E. and Mohan, A. T.},
  journal = {Phys. Rev. C},
  volume = {106},
  issue = {2},
  pages = {L021301},
  numpages = {6},
  year = {2022},
  month = {Aug},
  publisher = {American Physical Society},
  doi = {10.1103/PhysRevC.106.L021301},
  url = {https://link.aps.org/doi/10.1103/PhysRevC.106.L021301}
}

@article{PhysRevC.106.014305,
  title = {Nuclear masses learned from a probabilistic neural network},
  author = {Lovell, A. E. and Mohan, A. T. and Sprouse, T. M. and Mumpower, M. R.},
  journal = {Phys. Rev. C},
  volume = {106},
  issue = {1},
  pages = {014305},
  numpages = {9},
  year = {2022},
  month = {Jul},
  publisher = {American Physical Society},
  doi = {10.1103/PhysRevC.106.014305},
  url = {https://link.aps.org/doi/10.1103/PhysRevC.106.014305}
}

@Inbook{Arnold2009,
editor="Givental, Alexander B.
and Khesin, Boris A.
and Marsden, Jerrold E.
and Varchenko, Alexander N.
and Vassiliev, Victor A.
and Viro, Oleg Ya.
and Zakalyukin, Vladimir M.",
title="On the representation of functions of several variables as a superposition of functions of a smaller number of variables",
bookTitle="Collected Works: Representations of Functions, Celestial Mechanics and KAM Theory, 1957--1965",
year="2009",
publisher="Springer Berlin Heidelberg",
address="Berlin, Heidelberg",
pages="25--46",
isbn="978-3-642-01742-1",
doi="10.1007/978-3-642-01742-1_5",
url="https://doi.org/10.1007/978-3-642-01742-1_5",
author="V. I. Arnol'd"
}

@article{BETHE1936,
  title = {Nuclear Physics A. Stationary States of Nuclei},
  author = {Bethe, H. A. and Bacher, R. F.},
  journal = {Rev. Mod. Phys.},
  volume = {8},
  issue = {2},
  pages = {82--229},
  numpages = {0},
  year = {1936},
  month = {Apr},
  publisher = {American Physical Society},
  doi = {10.1103/RevModPhys.8.82},
  url = {https://link.aps.org/doi/10.1103/RevModPhys.8.82}
}

@article{weizsacker1935theorie,
  title={Zur theorie der kernmassen},
  author={Weizs{\"a}cker, CF v},
  journal={Z. Phys.},
  volume={96},
  number={7},
  pages={431--458},
  year={1935},
  publisher={Springer-Verlag Berlin/Heidelberg}
}

@article{RevModPhys.81.1773,
  title = {Modern theory of nuclear forces},
  author = {Epelbaum, E. and Hammer, H.-W. and Mei\ss{}ner, Ulf-G.},
  journal = {Rev. Mod. Phys.},
  volume = {81},
  issue = {4},
  pages = {1773--1825},
  numpages = {0},
  year = {2009},
  month = {Dec},
  publisher = {American Physical Society},
  doi = {10.1103/RevModPhys.81.1773},
  url = {https://link.aps.org/doi/10.1103/RevModPhys.81.1773}
}

@article{Hammer20,
  title = {Nuclear effective field theory: Status and perspectives},
  author = {Hammer, H.-W. and K\"onig, Sebastian and van Kolck, U.},
  journal = {Rev. Mod. Phys.},
  volume = {92},
  issue = {2},
  pages = {025004},
  numpages = {66},
  year = {2020},
  month = {Jun},
  publisher = {American Physical Society},
  doi = {10.1103/RevModPhys.92.025004},
  url = {https://link.aps.org/doi/10.1103/RevModPhys.92.025004}
}

@article{hadizadeh2020three,
  title={A three-dimensional momentum-space calculation of three-body bound state in a relativistic Faddeev scheme},
  author={Hadizadeh, MR and Radin, M and Mohseni, K},
  journal={Sci. Rep.},
  volume={10},
  number={1},
  pages={1949},
  year={2020},
  publisher={Nature Publishing Group UK London},
  doi={https://doi.org/10.1038/s41598-020-58577-4},
  url={https://www.nature.com/articles/s41598-020-58577-4}

}

@article{mean-field,
  title = {First Gogny-Hartree-Fock-Bogoliubov Nuclear Mass Model},
  author = {Goriely, S. and Hilaire, S. and Girod, M. and P\'eru, S.},
  journal = {Phys. Rev. Lett.},
  volume = {102},
  issue = {24},
  pages = {242501},
  numpages = {4},
  year = {2009},
  month = {Jun},
  publisher = {American Physical Society},
  doi = {10.1103/PhysRevLett.102.242501},
  url = {https://link.aps.org/doi/10.1103/PhysRevLett.102.242501}
}

@article{MYERS1996141,
title = {Nuclear properties according to the Thomas-Fermi model},
journal = {Nucl. Phys. A},
volume = {601},
number = {2},
pages = {141-167},
year = {1996},
issn = {0375-9474},
doi = {https://doi.org/10.1016/0375-9474(95)00509-9},
url = {https://www.sciencedirect.com/science/article/pii/0375947495005099},
author = {W.D. Myers and W.J. Swiatecki}
}

@article{ARNOULD200797,
title = {The r-process of stellar nucleosynthesis: Astrophysics and nuclear physics achievements and mysteries},
journal = {Phys. Rep.},
volume = {450},
number = {4},
pages = {97-213},
year = {2007},
issn = {0370-1573},
doi = {https://doi.org/10.1016/j.physrep.2007.06.002},
url = {https://www.sciencedirect.com/science/article/pii/S0370157307002438},
author = {M. Arnould and S. Goriely and K. Takahashi}
}

@article{yuan2024reliable,
  title={Reliable calculations of nuclear binding energies by the Gaussian process of machine learning},
  author={Yuan, Zi-Yi and Bai, Dong and Wang, Zhen and Ren, Zhong-Zhou},
  journal={Nucl. Sci. Tech.},
  volume={35},
  number={6},
  pages={105},
  year={2024},
  publisher={Springer}
}

@article{PhysRevC.104.014315,
  title = {Improved naive Bayesian probability classifier in predictions of nuclear mass},
  author = {Liu, Yifan and Su, Chen and Liu, Jian and Danielewicz, Pawel and Xu, Chang and Ren, Zhongzhou},
  journal = {Phys. Rev. C},
  volume = {104},
  issue = {1},
  pages = {014315},
  numpages = {9},
  year = {2021},
  month = {Jul},
  publisher = {American Physical Society},
  doi = {10.1103/PhysRevC.104.014315},
  url = {https://link.aps.org/doi/10.1103/PhysRevC.104.014315}
}

@article{PhysRevC.98.034318,
  title = {Bayesian approach to model-based extrapolation of nuclear observables},
  author = {Neufcourt, L\'eo and Cao, Yuchen and Nazarewicz, Witold and Viens, Frederi},
  journal = {Phys. Rev. C},
  volume = {98},
  issue = {3},
  pages = {034318},
  numpages = {17},
  year = {2018},
  month = {Sep},
  publisher = {American Physical Society},
  doi = {10.1103/PhysRevC.98.034318},
  url = {https://link.aps.org/doi/10.1103/PhysRevC.98.034318}
}

@article{PhysRevC.105.064306,
  title = {Deep learning approach to nuclear masses and $\ensuremath{\alpha}$-decay half-lives},
  author = {Li, Chen-Qi and Tong, Chao-Nan and Du, Hong-Jing and Pang, Long-Gang},
  journal = {Phys. Rev. C},
  volume = {105},
  issue = {6},
  pages = {064306},
  numpages = {10},
  year = {2022},
  month = {Jun},
  publisher = {American Physical Society},
  doi = {10.1103/PhysRevC.105.064306},
  url = {https://link.aps.org/doi/10.1103/PhysRevC.105.064306}
}

@article{xraySchatz_2017,
doi = {10.3847/1538-4357/aa7de9},
url = {https://dx.doi.org/10.3847/1538-4357/aa7de9},
year = {2017},
month = {aug},
publisher = {The American Astronomical Society},
volume = {844},
number = {2},
pages = {139},
author = {H. Schatz and W.-J. Ong},
title = {Dependence of X-Ray Burst Models on Nuclear Masses},
journal = {Astrophys. J.}
}

@article{PhysRevLett.114.202501,
  title = {Probing the $N=32$ Shell Closure below the Magic Proton Number $Z=20$: Mass Measurements of the Exotic Isotopes $^{52,53}\mathrm{K}$},
  author = {Rosenbusch, M. and Ascher, P. and Atanasov, D. and Barbieri, C. and Beck, D. and Blaum, K. and Borgmann, Ch. and Breitenfeldt, M. and Cakirli, R. B. and Cipollone, A. and George, S. and Herfurth, F. and Kowalska, M. and Kreim, S. and Lunney, D. and Manea, V. and Navr\'atil, P. and Neidherr, D. and Schweikhard, L. and Som\`a, V. and Stanja, J. and Wienholtz, F. and Wolf, R. N. and Zuber, K.},
  journal = {Phys. Rev. Lett.},
  volume = {114},
  issue = {20},
  pages = {202501},
  numpages = {6},
  year = {2015},
  month = {May},
  publisher = {American Physical Society},
  doi = {10.1103/PhysRevLett.114.202501},
  url = {https://link.aps.org/doi/10.1103/PhysRevLett.114.202501}
}

@article{PhysRevC.90.024320,
  title = {Systematic study of shell gaps in nuclei},
  author = {Mo, Qiuhong and Liu, Min and Wang, Ning},
  journal = {Phys. Rev. C},
  volume = {90},
  issue = {2},
  pages = {024320},
  numpages = {8},
  year = {2014},
  month = {Aug},
  publisher = {American Physical Society},
  doi = {10.1103/PhysRevC.90.024320},
  url = {https://link.aps.org/doi/10.1103/PhysRevC.90.024320}
}

@article{RevModPhys.75.1021,
  title = {Recent trends in the determination of nuclear masses},
  author = {Lunney, D. and Pearson, J. M. and Thibault, C.},
  journal = {Rev. Mod. Phys.},
  volume = {75},
  issue = {3},
  pages = {1021--1082},
  numpages = {0},
  year = {2003},
  month = {Aug},
  publisher = {American Physical Society},
  doi = {10.1103/RevModPhys.75.1021},
  url = {https://link.aps.org/doi/10.1103/RevModPhys.75.1021}
}

@article{PhysRevC.106.L021303,
  title = {Nuclear mass predictions with machine learning reaching the accuracy required by $r$-process studies},
  author = {Niu, Z. M. and Liang, H. Z.},
  journal = {Phys. Rev. C},
  volume = {106},
  issue = {2},
  pages = {L021303},
  numpages = {6},
  year = {2022},
  month = {Aug},
  publisher = {American Physical Society},
  doi = {10.1103/PhysRevC.106.L021303},
  url = {https://link.aps.org/doi/10.1103/PhysRevC.106.L021303}
}

@article{kfold1,
author = {David M. Allen},
title = {The Relationship Between Variable Selection and Data Agumentation and a Method for Prediction},
journal = {Technometrics},
volume = {16},
number = {1},
pages = {125--127},
year = {1974},
publisher = {Taylor \& Francis},
doi = {10.1080/00401706.1974.10489157},
}

@article{kfold2,
    author = {Stone, M.},
    title = "{Cross-Validatory Choice and Assessment of Statistical Predictions}",
    journal = {J. R. Stat. Soc. B},
    volume = {36},
    number = {2},
    pages = {111-133},
    year = {2018},
    month = {12},
    issn = {0035-9246},
    doi = {10.1111/j.2517-6161.1974.tb00994.x},
    url = {https://doi.org/10.1111/j.2517-6161.1974.tb00994.x},
}

@article{kfold3,
    author = {Stone, M.},
    title = "{An Asymptotic Equivalence of Choice of Model by Cross-Validation and Akaike's Criterion}",
    journal = {J. R. Stat. Soc. B},
    volume = {39},
    number = {1},
    pages = {44-47},
    year = {2018},
    month = {12},
    issn = {0035-9246},
    doi = {10.1111/j.2517-6161.1977.tb01603.x},
    url = {https://doi.org/10.1111/j.2517-6161.1977.tb01603.x},
}

@misc{munoz2024discoveringnuclearmodelssymbolic,
      title={Discovering Nuclear Models from Symbolic Machine Learning}, 
      author={Jose M. Munoz and Silviu M. Udrescu and Ronald F. Garcia Ruiz},
      year={2024},
      eprint={2404.11477},
      archivePrefix={arXiv},
      primaryClass={nucl-th},
      url={https://arxiv.org/abs/2404.11477}, 
}

@article{MOLLER20161,
title = {Nuclear ground-state masses and deformations: FRDM(2012)},
journal = {At. Data Nucl. Data Tables},
volume = {109-110},
pages = {1-204},
year = {2016},
issn = {0092-640X},
doi = {https://doi.org/10.1016/j.adt.2015.10.002},
url = {https://www.sciencedirect.com/science/article/pii/S0092640X1600005X},
author = {P. Möller and A.J. Sierk and T. Ichikawa and H. Sagawa}
}

@misc{warin2024p1kaneffectivekolmogorovarnold,
      title={P1-KAN an effective Kolmogorov Arnold Network for function approximation}, 
      author={Xavier Warin},
      year={2024},
      eprint={2410.03801},
      archivePrefix={arXiv},
      primaryClass={cs.LG},
      url={https://arxiv.org/abs/2410.03801}, 
}

@misc{aghaei2024fkan,
      title={fKAN: Fractional Kolmogorov-Arnold Networks with trainable Jacobi basis functions},
      author={Alireza Afzal Aghaei},
      year={2024},
      eprint={2406.07456},
      archivePrefix={arXiv},
      primaryClass={cs.LG},
      url={https://arxiv.org/abs/2406.07456}
}

@misc{aghaei2024rkanrationalkolmogorovarnoldnetworks,
      title={rKAN: Rational Kolmogorov-Arnold Networks}, 
      author={Alireza Afzal Aghaei},
      year={2024},
      eprint={2406.14495},
      archivePrefix={arXiv},
      primaryClass={cs.LG},
      url={https://arxiv.org/abs/2406.14495}, 
}

@article{PhysRevC.93.034337,
  title = {Further explorations of Skyrme-Hartree-Fock-Bogoliubov mass formulas. XVI. Inclusion of self-energy effects in pairing},
  author = {Goriely, S. and Chamel, N. and Pearson, J. M.},
  journal = {Phys. Rev. C},
  volume = {93},
  issue = {3},
  pages = {034337},
  numpages = {11},
  year = {2016},
  month = {Mar},
  publisher = {American Physical Society},
  doi = {10.1103/PhysRevC.93.034337},
  url = {https://link.aps.org/doi/10.1103/PhysRevC.93.034337}
}

@misc{giroux2024uncertaintyquantificationbayesianhigher,
      title={Uncertainty Quantification with Bayesian Higher Order ReLU KANs}, 
      author={James Giroux and Cristiano Fanelli},
      year={2024},
      eprint={2410.01687},
      archivePrefix={arXiv},
      primaryClass={cs.LG},
      url={https://arxiv.org/abs/2410.01687}, 
}

@article{Menditto2007UnderstandingTM,
  title={Understanding the meaning of accuracy, trueness and precision},
  author={Antonio Menditto and Marina Patriarca and Bertil Magnusson},
  journal={Accredit. Qual. Assur.},
  year={2007},
  volume={12},
  pages={45-47},
  url={https://api.semanticscholar.org/CorpusID:109596162}
}

@inproceedings{
liu2025kan,
title={{KAN}: Kolmogorov{\textendash}Arnold Networks},
author={Ziming Liu and Yixuan Wang and Sachin Vaidya and Fabian Ruehle and James Halverson and Marin Solja{\v c}i{\'c} and Thomas Y. Hou and Max Tegmark},
booktitle={The Thirteenth International Conference on Learning Representations},
year={2025},
url={https://openreview.net/forum?id=Ozo7qJ5vZi}
}

@article{KAM1957,
author = {Kolmogorov, A. N.},
journal = {Dokl. Akad. Nauk SSSR},
volume = {114},
pages = {953},
year = {1957},
}

@book{19531603584,
author = {Fisher, R. A. and Yates, F.},
title = {Statistical Tables for Biological, Agricultural and Medical Research},
year = {1953},
edition = {4th},
pages = {26--27},
publisher = {Oliver \& Boyd},
address = {Edinburgh},
}

@ARTICLE{10.3389/frai.2024.1462952,

AUTHOR={Reinhardt, Eric  and Ramakrishnan, Dinesh  and Gleyzer, Sergei },

TITLE={SineKAN: Kolmogorov-Arnold Networks using sinusoidal activation functions},

JOURNAL={Front. Artif. Intell.},

VOLUME={7},

PAGES={1462952},

YEAR={2025},

URL={https://www.frontiersin.org/journals/artificial-intelligence/articles/10.3389/frai.2024.1462952},

DOI={10.3389/frai.2024.1462952},

ISSN={2624-8212},

}

\end{document}